\documentclass{PoS}

\usepackage{amsmath}
\usepackage{subfig}
\usepackage{cite}
\usepackage{url}

\newcommand{\ncteq}{{\tt nCTEQ}}
\newcommand{\ncteqfit}{{\tt nCTEQ15}}
\newcommand{\ncteqfitmod}{{\tt nCTEQ15-mod}}
\newcommand{\ncteqnp}{{\tt nCTEQ15-np}}

\vbox{
\null \vspace{0.3in}
\hbox{LPSC-15-181}
}

\title{nCTEQ15 -- Global analysis of nuclear parton distributions with uncertainties}

\ShortTitle{nCTEQ15 global analysis}

\author{\speaker{A. Kusina}$^a$, K.~Kova\v{r}\'{\i}k$^b$, T.~Je\v{z}o$^c$,
        D.~B.~Clark$^d$, C.~Keppel$^e$, F.~Lyonnet$^d$, J.G.~Morf{\'i}n$^f$
        F.~I.~Olness$^d$, J.F.~Owens$^g$, I.~Schienbein$^a$, J.~Y.~Yu$^d$\\
        \llap{$^a$}Laboratoire de Physique Subatomique et de Cosmologie,
                   Universit\'e Grenoble-Alpes, CNRS/IN2P3,
                   53 avenue des Martyrs, 38026 Grenoble, France\\
        \llap{$^b$}Institut f{\"u}r Theoretische Physik, Westf{\"a}lische Wilhelms-Universit{\"a}t M{\"u}nster,
                   Wilhelm-Klemm-Stra{\ss}e 9, D-48149 M{\"u}nster, Germany\\
        \llap{$^c$}Universit\`a di Milano-Bicocca and INFN, Sezione di Milano-Bicocca,\\
                   Piazza della Scienza 3, 20126 Milano, Italy\\
        \llap{$^d$}Southern Methodist University, Dallas, TX 75275, USA\\
        \llap{$^e$}Thomas Jefferson National Accelerator Facility, Newport News, VA, 23606, USA\\
        \llap{$^f$}Fermi National Accelerator Laboratory, Batavia, Illinois 60510, USA\\
        \llap{$^g$}Department of Physics, Florida State University, Tallahassee, Florida 32306-4350, USA\\
        E-mail: \email{kusina@lpsc.in2p3.fr}, \email{karol.kovarik@uni-muenster.de},
                \email{tomas.jezo@mib.infn.it}, \email{dbclark@smu.edu}, \email{keppel@jlab.org},
                \email{flyonnet@smu.edu}, \email{morfin@fnal.gov}, \email{olness@smu.edu},
                \email{owens@hep.fsu.edu}, \email{ingo.schienbein@lpsc.in2p3.fr}, \email{yu@physics.smu.edu}}

\abstract{We present the first official release of the nCTEQ nuclear parton distribution
functions with errors. The main addition to the previous nCTEQ PDFs is the introduction of
PDF uncertainties based on the Hessian method. Another important addition is the inclusion
of pion production data from RHIC that give us a handle on constraining the gluon PDF.
This contribution summarizes our results from~\cite{Kovarik:2015cma} and concentrates on the comparison
with other groups providing nuclear parton distributions.}

\FullConference{The XXIII International Workshop on Deep Inelastic Scattering and Related Subjects\\
		April 27 - May 1, 2015\\
                Southern Methodist University\\
		Dallas, Texas 75275}

\begin{document}

\section{Introduction}
Nucleons and nuclei can be described using the language of parton distribution
functions (PDFs) which is based on factorization
theorems~\cite{Collins:1985ue,Bodwin:1984hc,Qiu:2003cg,Qiu:2002mh}.
The case of a free proton is extremely  well studied. Several global analyses of
free proton PDFs, based on an ever growing set of precise experimental data and
on next-to-next-to-leading order theoretical predictions, are regularly updated
and maintained~\cite{Martin:2009iq,Gao:2013xoa,Ball:2012cx,Alekhin:2013nda,Owens:2012bv}.
%
The structure of a nucleus can be effectively parametrized in terms of protons bound
inside a nucleus and described by nuclear PDFs (nPDFs). These nPDFs contain effects
on proton structure coming from the strong interactions between
the nucleons in a nucleus. Similarly to the PDFs of free protons, nuclear PDFs are
obtained by fitting experimental data including deep inelastic scattering on nuclei
and nuclear collision experiments. Moreover, as the nuclear effects are clearly dependent
on the number of nucleons, experimental data from scattering on multiple nuclei must
be considered. In contrast to the free proton PDFs where quark distributions for most
flavors together with the gluon distribution are reliably determined over a large
kinematic range,  nuclear PDFs precision is not comparable due to the lack of accuracy
of the current relevant data. In addition, the non-trivial dependence of nuclear effects
on the number of nucleons requires a large data set involving several different nuclei.
Nevertheless, nuclear PDFs are a crucial ingredient in predictions for high energy
collisions involving nuclear targets, such as the lead collisions performed at the LHC.

In this contribution we present the new \ncteq\ nuclear PDFs that were recently
released~\cite{Kovarik:2015cma} and compare them with analyses from other groups providing
nPDFs~\cite{Hirai:2007sx,Eskola:2009uj,deFlorian:2011fp}. All the details of the analysis
can be found in ref.~\cite{Kovarik:2015cma} here we will mostly concentrate on the differences
with other nPDFs.

\section{\ncteq\ global analysis}
In the presented \ncteq\ analysis we use mostly charged lepton deep inelastic scattering
(DIS) and Drell-Yan process (DY) data that provide respectively 616 and 92 data points.
Additionally we include pion production data from RHIC (32 data points) that have potential
to constrain the gluon PDF. To better asses the impact of the pion data on our analysis
two fits are discussed: (i) the main \ncteqfit\ fit using all aforementioned data,
and (ii) \ncteqnp\ fit which does not include the pion data.
The framework of the current analysis, including parameterization, fitting procedure and
precise prescription for the Hessian method used to estimate PDF uncertainties is
defined in ref.~\cite{Kovarik:2015cma} and we refer reader to this paper for details.

In both presented fits we use 16 free parameters to describe the nPDFs, that comprise
7 gluon, 4 $u$-valence, 3 $d$-valence and 2 $\bar{d}+\bar{u}$ parameters.
In addition, in the \ncteqfit\ case the normalization of the pion data sets is fitted
which adds two more free parameters.
%
Both our fits, \ncteqfit\ and \ncteqnp\ describe the data very well.
Indeed, the quality of the fits as measured by the values of the $\chi^2/$dof
(0.85 and 0.87 for the \ncteqfit\ and \ncteqnp\ fits respectively), confirms it.
%
\begin{figure*}[th]
\centering{}
\includegraphics[clip,width=0.48\textwidth]{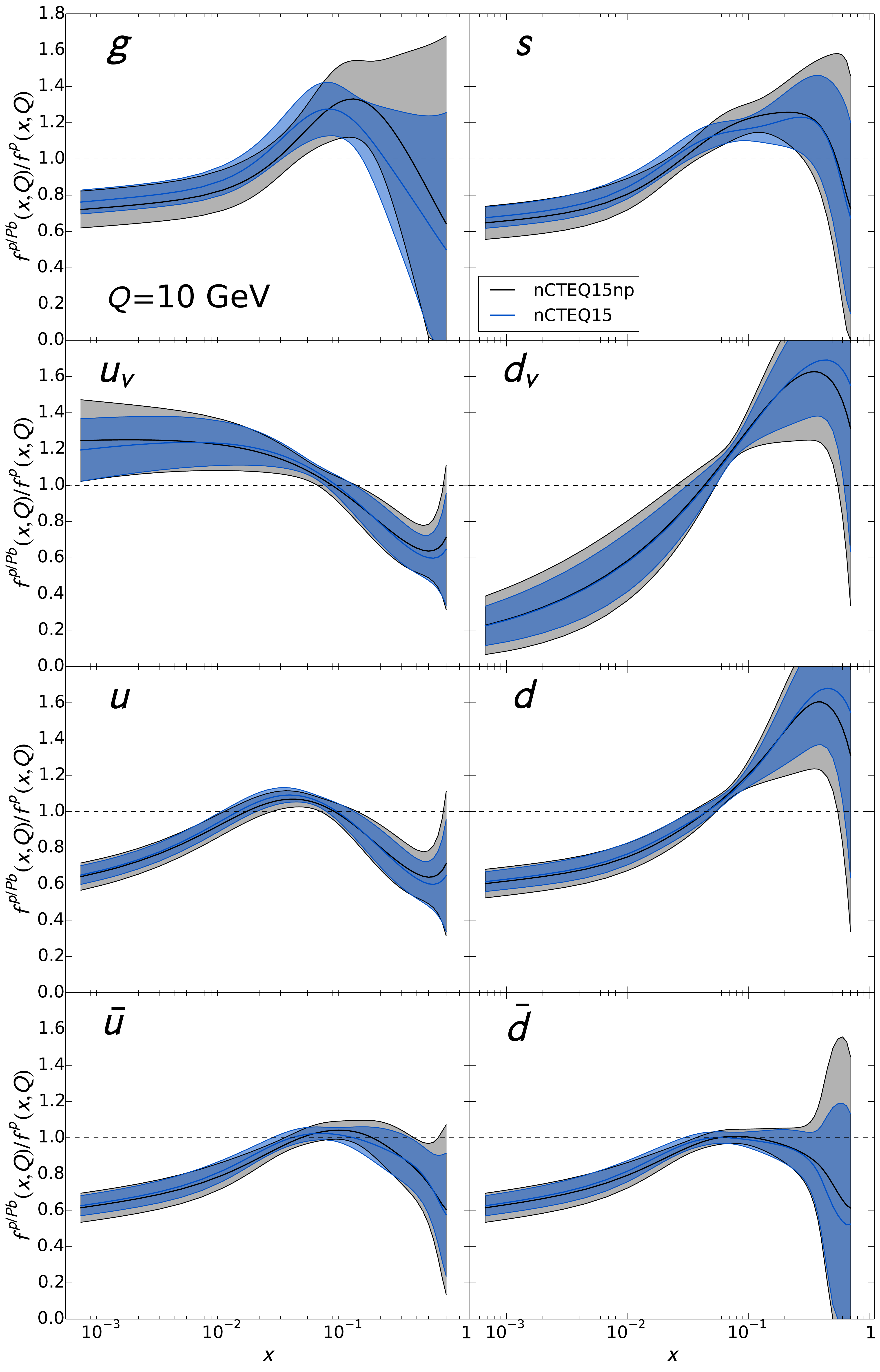}
\quad{}
\includegraphics[width=0.48\textwidth]{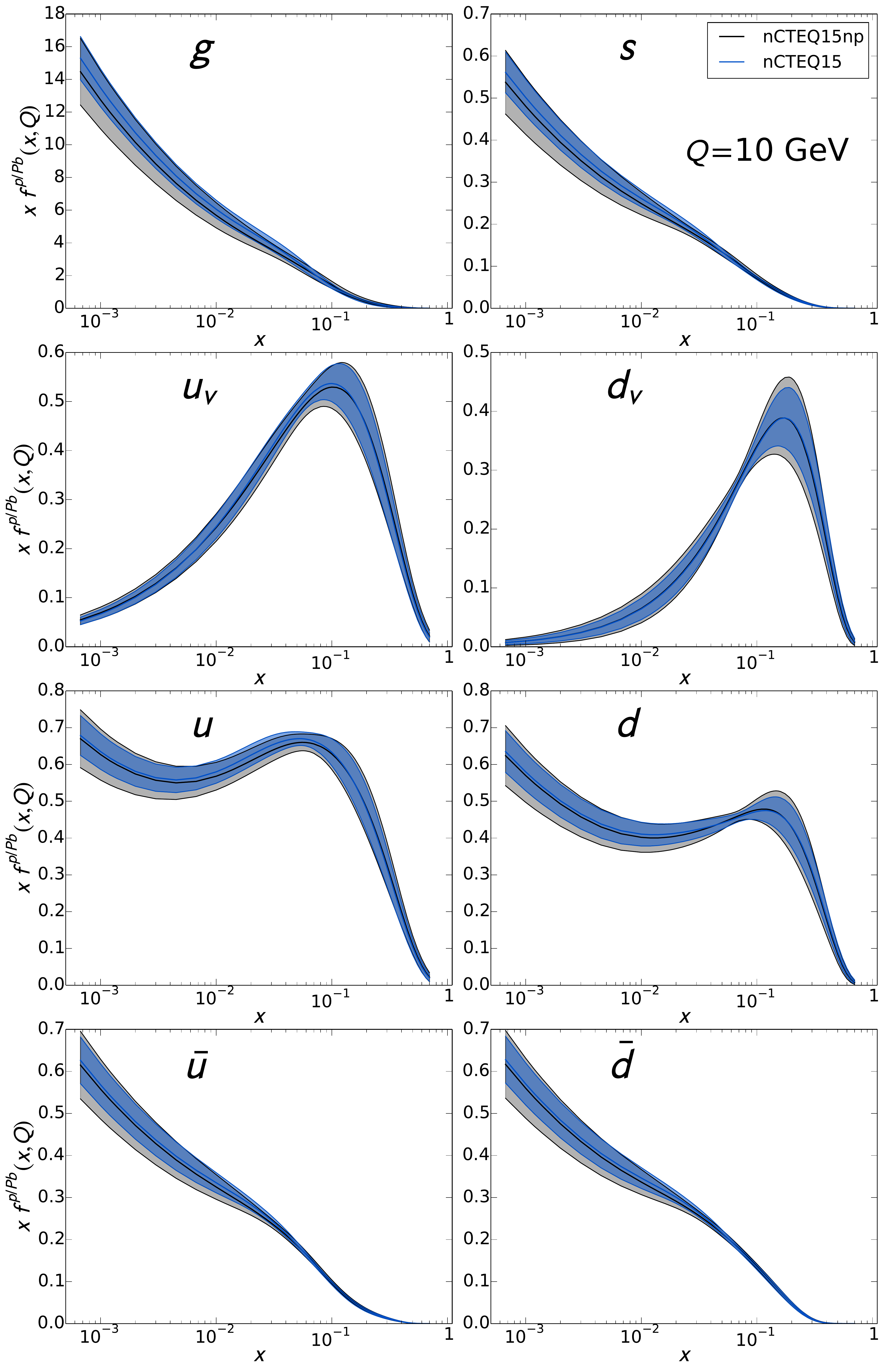}
\caption{Comparison of the \ncteqfit\ fit (blue)
with the
\ncteqnp\ fit without pion data (gray).
On the left we show nuclear modification factors defined as ratios of proton PDFs
bound in lead to the corresponding free proton PDFs, and on the right we show the
actual bound proton PDFs for lead. In both cases scale is equal to $Q=10$ GeV.}
\label{fig:e11NLOa-vs-decut4p16}
\end{figure*}
%
Figure~\ref{fig:e11NLOa-vs-decut4p16} shows the bound proton PDFs resulting from the two fits.
The nuclear correction factors (left panel) clearly show that the pion data impact
the gluon distribution, and to a lesser extent the $u_v$, $d_v$ and $s$ PDFs.
%
Also, the inclusion of the pion data decreases the lead gluon PDF at larger $x$
($\gtrsim 10^{-1}$), and increases it at smaller $x$ whereas the error bands are
reduced in the intermediate to larger $x$ range.
For most of the other PDF flavors, the change in the central value is minimal (except for
a few cases at high-$x$ where the magnitude of the PDF is small).  For these other PDFs,
the inclusion of the pion data generally decreases the size of the error band.

\begin{figure*}[th]
\begin{center}
\includegraphics[width=0.8\textwidth]{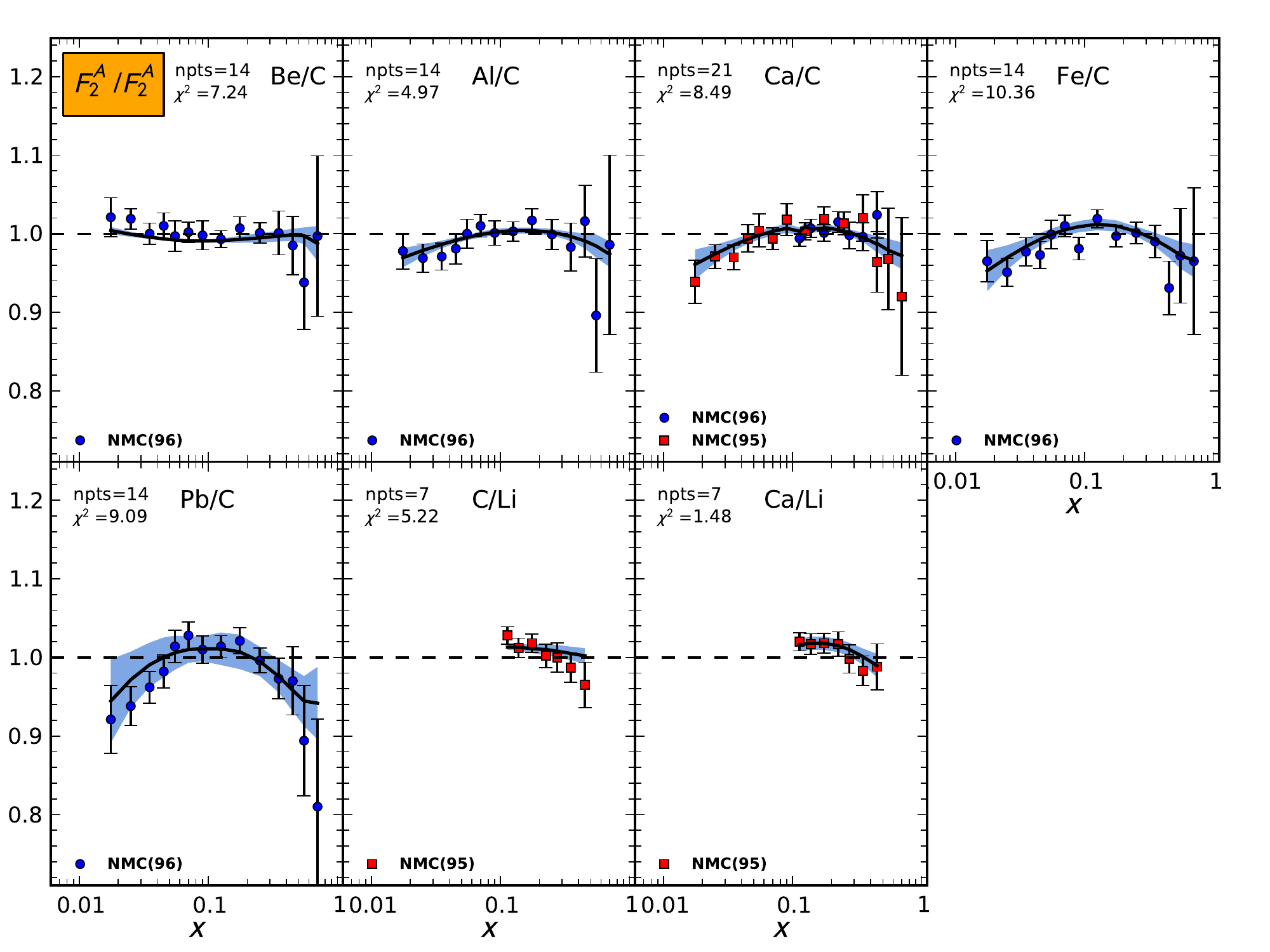}
\caption{Comparison of the \ncteqfit\ NLO theory predictions for $R=F_2^A/F_2^{A'}$ 
with  nuclear target data. 
The bands show the  uncertainty from the nuclear PDFs.}
\label{fig:F2ratioVsX}
\end{center}
\end{figure*}

\begin{figure*}[th]
\begin{center}
\subfloat[]
{
\includegraphics[width=0.47\textwidth]{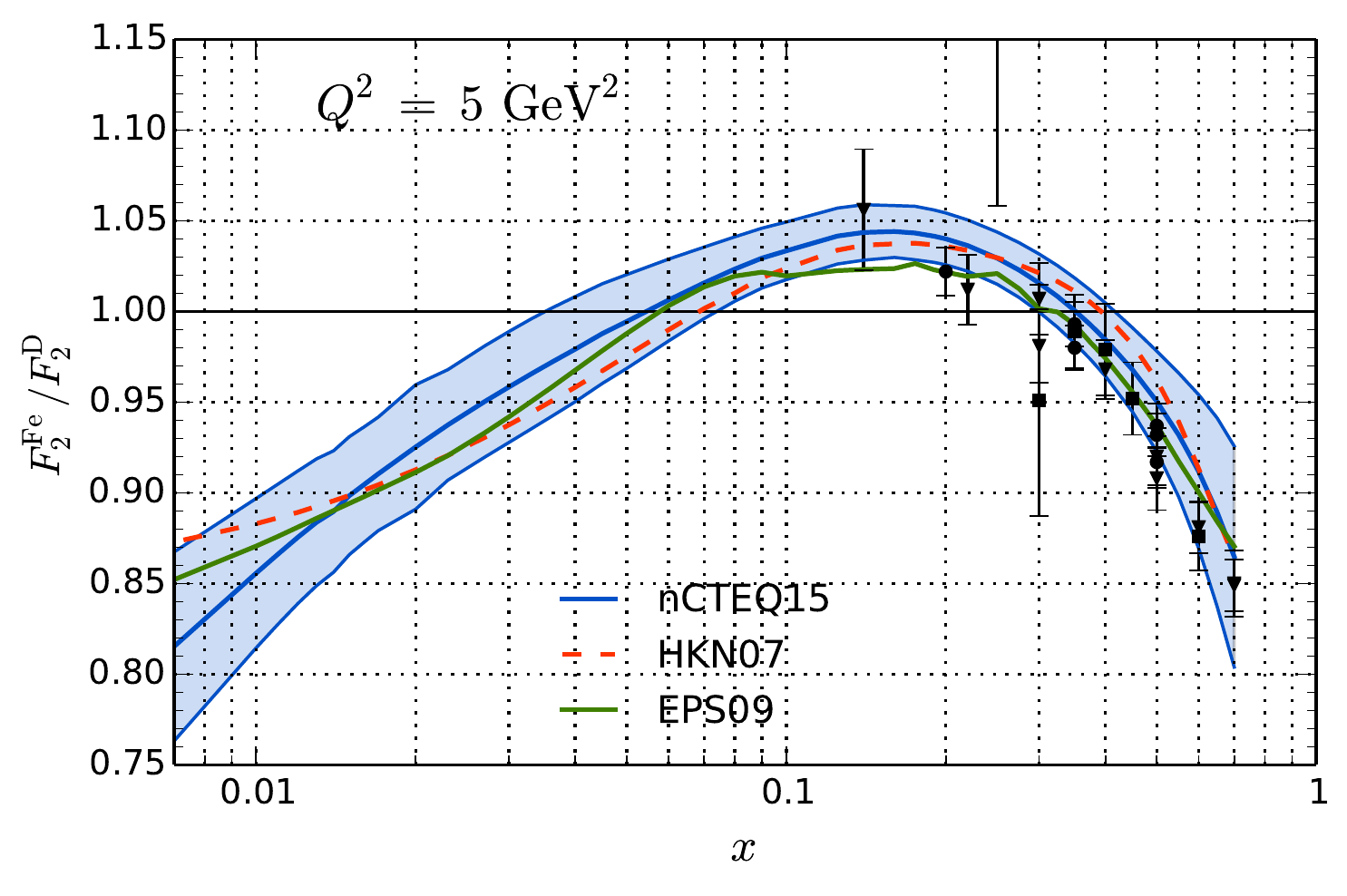}
\label{subfig:F2Fe_Q5}
}
\subfloat[]
{
\includegraphics[width=0.47\textwidth]{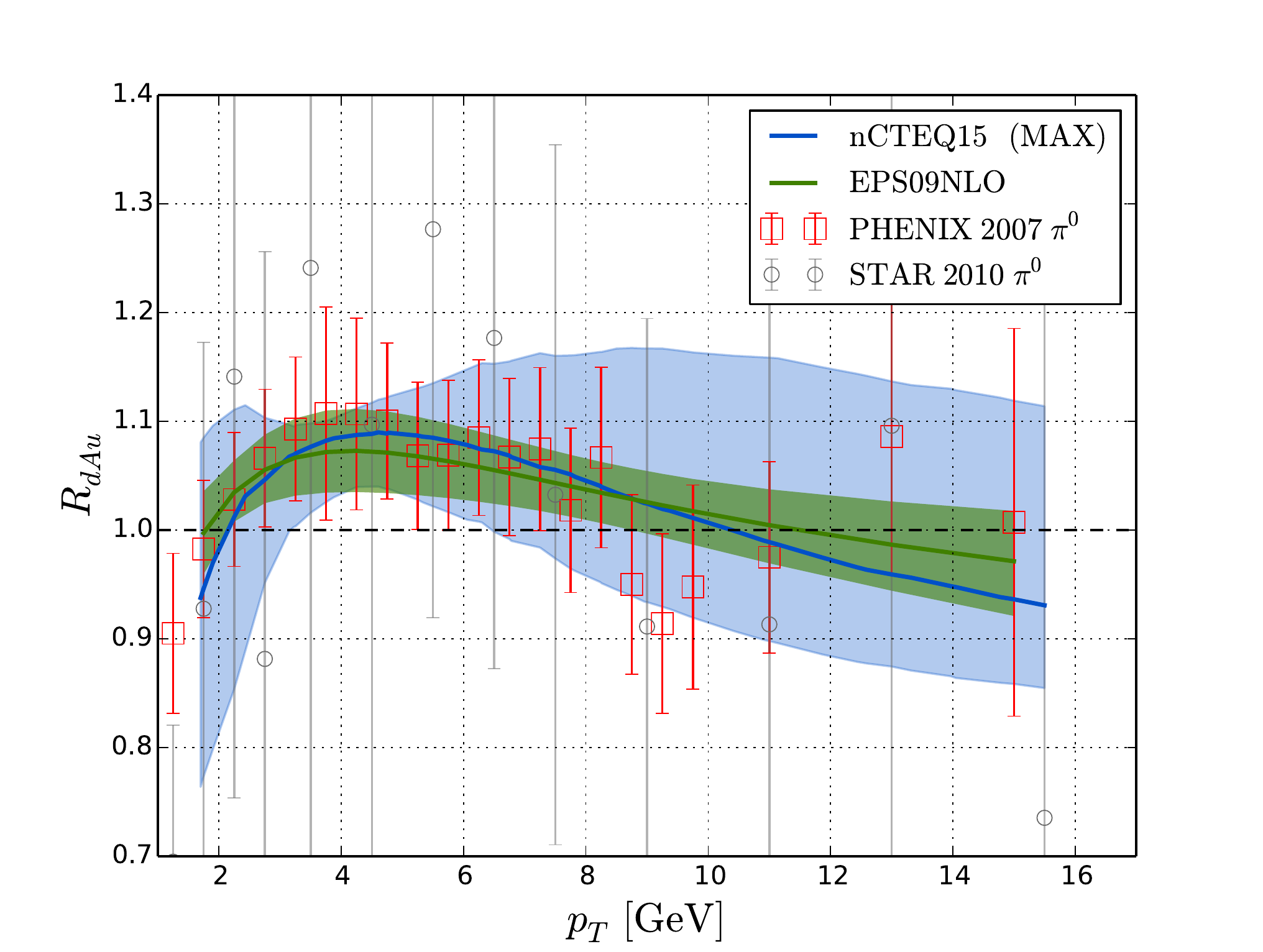}
\label{subfig:RdAu}
}
\caption{
(a) Ratio of the $F_2$ structure functions for iron and deuteron calculated
with the \ncteqfit\ fit at $Q^2=5~\text{GeV}^2$ overlaid with fitted
data~\cite{Bari:1985ga,Benvenuti:1987az,Bodek:1983ec,Gomez:1993ri,Dasu:1993vk}
and results from EPS09 and HKN07.
(b) Comparison of the \ncteqfit\ and EPS09 fits with the 
PHENIX~\cite{Adler:2006wg} and STAR~\cite{Abelev:2009hx} data 
for the ratio $R_{\text{dAu}}^{\pi} = \frac{\tfrac{1}{2A}d^2\sigma_{\pi}^{\text{dAu}}/dp_T dy}{d^2\sigma_{\pi}^{\text{pp}}/dp_T dy}$.
}
\label{fig:F2-RdAu}
\end{center}
\end{figure*}

The description of the fitted data by the \ncteqfit\ fit can be seen in
Fig.~\ref{fig:F2ratioVsX} where we display ratios of $F_2$ structure functions for different
nuclei
as well as the corresponding data.
Finally, in Fig.\ref{fig:F2-RdAu}, we display the comparison for the ratios of:
$F_2$ structure functions for iron over deuteron (left),
and pion yields in $DAu$ over $pp$ collisions at RHIC (right).
In both cases the \ncteqfit\ fit provides a very good description of the
experimental data.
%
%
%


\section{Comparison with other nPDFs}

\begin{figure*}[th]
\centering{}
\includegraphics[clip,width=0.48\textwidth]{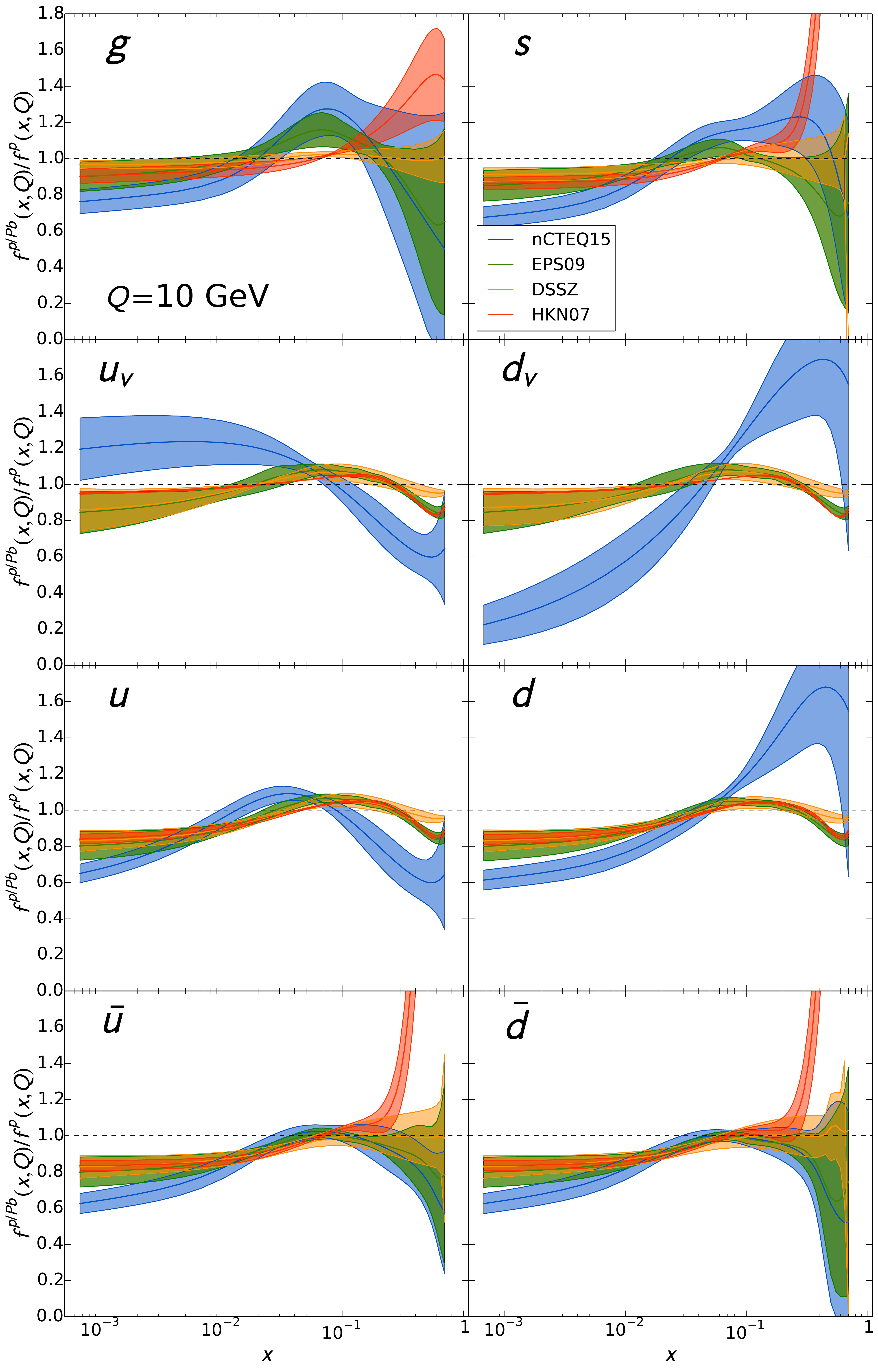}
\quad{}
\includegraphics[width=0.48\textwidth]{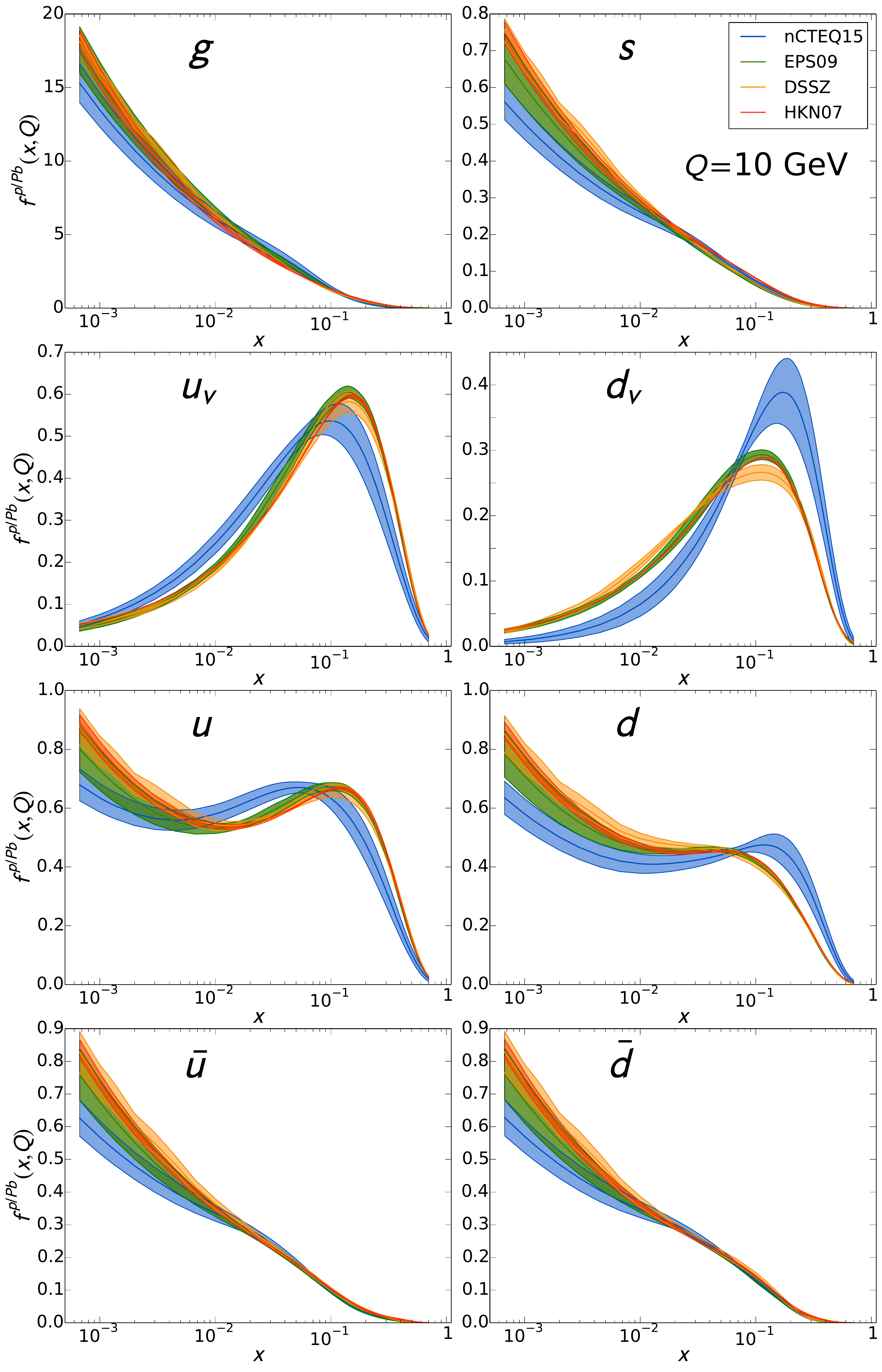}
\caption{
Comparison of  the \ncteqfit\ fit (blue)
with results from other groups:
EPS09~\cite{Eskola:2009uj} (green),
DSSZ~\cite{deFlorian:2011fp} (orange),
HKN07~\cite{Hirai:2007sx} (red).
The left panel shows nuclear modification factors for lead,
 and the right panel the actual
PDFs of a proton bound in \emph{lead}. 
The scale is $Q=10$~GeV.
}
\label{fig:compar_PDFs_Q10}
\end{figure*}

We now compare the \ncteqfit\ PDFs with other recent nuclear parton distributions
in the literature, in particular 
HKN07~\cite{Hirai:2007sx},
EPS09~\cite{Eskola:2009uj}, and
DSSZ~\cite{deFlorian:2011fp}.
Our data set selection and technical aspects of our analysis are
closer to that of EPS09 on which we focus our comparison in the following.
Our results for the nuclear modifications of the lead PDFs as well as the bound proton lead PDFs
themselves are shown in Fig.~\ref{fig:compar_PDFs_Q10} for different flavors at
the scale $Q=10$~GeV.
%

For most flavors, $\bar{u}$, $\bar{d}$, $s$ and $g$, there is a
reasonable agreement between predictions from different groups.
In particular, for the gluon, there is a larger spread in the predictions form
the various PDF sets;
we can see a distinct shape predicted by the \ncteqfit\ and EPS09 fits
whereas HKN07 and DSSZ have similar, much flatter behavior in the small to
intermediate $x$ region and deviates form each other in the higher $x$ region;
however, all these differences are nearly contained within the PDF uncertainty bands.


On the other hand, examining the $u$- and $d$-valence distributions one can see a very
different pattern.  Three of the PDF sets, HKN07, EPS09 and DSSZ,
are very similar across the $x$ range,
while the \ncteqfit\ set displays a marked
deviation from this behavior.%
    \footnote{This difference is significantly reduced when we consider full nuclear PDFs,
    $f^{A}=Z/A f^{p/A} + (A-Z)/Z f^{n/A}$, which are the quantities entering cross section
    calculations (see Fig.~25 in~\cite{Kovarik:2015cma}).}
At small-$x$ ($\lesssim 10^{-1}$) the $u_v$ is above the others while the $d_v$ is below;
and, for large-$x$ the exact opposite behavior is observed. This trend persists across
all $Q$ values. 
In essence, the {\em average} value of the $u_v$ and $d_v$ nuclear corrections are
comparable to the other groups, but individually the corrections are very different. 

\begin{figure*}[h!!!]
\centering{}
\includegraphics[clip=true,width=0.52\textwidth]{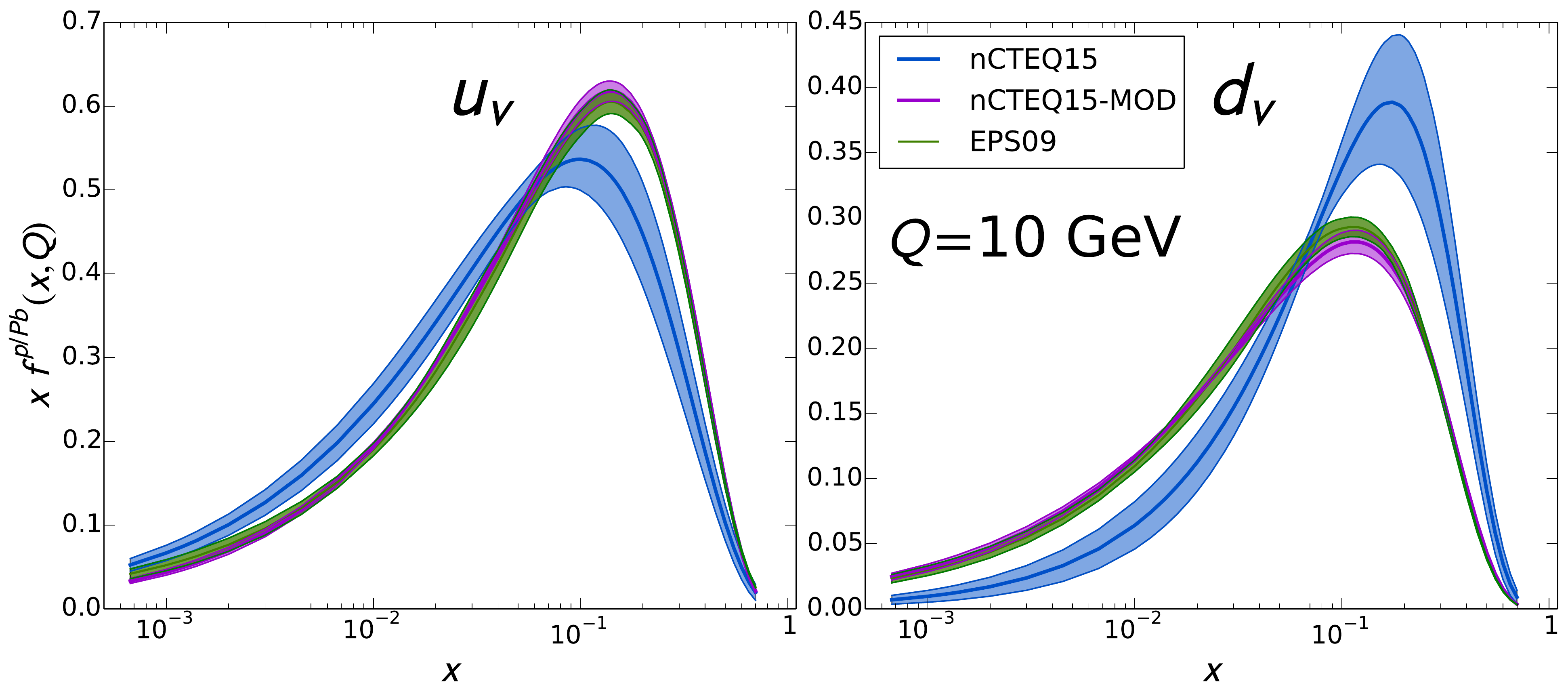}
\caption{
Bound proton $u_v$ and $d_v$ PDFs for lead for a modified \ncteqfitmod\ fit
using universal nuclear correction for both valence distributions (purple).
For comparison we show distributions for the \ncteqfit\ fit (blue) and EPS09 fit (green).
The scale is $Q=10$ GeV.
}
\label{fig:moduvdv}
\end{figure*}

This difference highlights an essential feature of the \ncteqfit\ fit; namely,
that the $u_v$ and $d_v$ are allowed to be independent, whereas other groups assume
the corresponding nuclear corrections to be identical. 
Certainly there is no physical motivation to assume the $u_v$ and $d_v$ nuclear corrections
to be universal; we exploit this freedom and fit the two PDFs independently.
To demonstrate that the difference in the shape of the valence distributions is not
an artifact of our methodology, 
but rather the result of the data and the physically motivated freedom in valence nuclear corrections,
we performed an additional fit, \ncteqfitmod,
in which the  $u_v$ and $d_v$ parameters have been tied together in order to obtained
a universal correction.
%
This result is shown in Fig.~\ref{fig:moduvdv}  where the \ncteqfitmod\ fit
aligns closely with the EPS09 prediction. The $\chi^2$ of the \ncteqfitmod\ fit increases
to 677 ($\chi^2/{\rm dof}=0.94$) as compared to 611 ($\chi^2/{\rm dof}=0.85$) for the
\ncteqfit\ fit. This is more than our tolerance criteria, $\Delta\chi^2=35$,
used for defining the error PDFs, and indicates that the data prefer this additional freedom.

Looking at Fig.~\ref{fig:F2-RdAu} it is also clear that the differences in the valence
distributions do not influence the ability to describe the currently available data.
We can see that observables in Fig.~\ref{fig:F2-RdAu} are very well described by both 
\ncteqfit\ and EPS09 fits despite the very different valence PDFs.%
    \footnote{The very notable difference of \ncteqfit\ and EPS09 error band in
    Fig.\ref{subfig:RdAu} is caused by the fact that EPS09 include pion data with weight
    of 20 (whereas we use weight of 1) which effectively shrinks the errors.}
This fact highlights the need for more data to constrain the nuclear PDFs.

\section{Summary}
We have presented the \ncteqfit\ set of nuclear PDFs including a set of error nPDFs
generated using the Hessian method. In addition to the standard DIS and DY data sets,
pion production data from RHIC were also included, providing an additional handle on
the gluon PDF.

We have compared the results of the \ncteqfit\ fit with nPDFs from HKN07, EPS09,
and DSSZ groups. While there are similarities between the \ncteqfit\ fit and the
other sets at a macro level, there are significant differences in the details.
In particular, the most notable and important difference is the treatment of the
$u$-valence and $d$-valence nuclear corrections. While other groups use a universal
nuclear correction for the valence distributions we believe that there is no physical
reason for doing so and we treat them as independent.
This additional freedom leads to an improved fit with significantly lower $\chi^2$
which in turn is preferred by the available data.
However, to clearly answer this question, more
data allowing the separation between $u$ and $d$ quarks will be needed.
One example of such a data is the $W/Z$ boson production in $pPb$ collisions at the LHC,
e.g.~\cite{Khachatryan:2015hha}.

Our PDFs are publicly available at the \ncteq\ website: \url{www.ncteq.org}.
We provide grids in the internal CTEQ PDS format (together with the corresponding interface)
and also in the new LHAPDF6 format~\cite{Buckley:2014ana}.

\bibliographystyle{utphys_spires}
\bibliography{biblio}

\providecommand{\href}[2]{#2}\begingroup\begin{thebibliography}{10}

\bibitem{Kovarik:2015cma}
{nCTEQ} Collaboration, K.~Kovarik {\em et al.},
\href{http://www.arXiv.org/abs/1509.00792}{{\tt 1509.00792}}.

\bibitem{Collins:1985ue}
J.~C. Collins, D.~E. Soper, and G.~F. Sterman, {\em Nucl.Phys.} {\bf B261}
  (1985)
104.

\bibitem{Bodwin:1984hc}
G.~T. Bodwin, {\em Phys.Rev.} {\bf D31} (1985)
2616.

\bibitem{Qiu:2003cg}
J.-w. Qiu,
\href{http://www.arXiv.org/abs/hep-ph/0305161}{{\tt hep-ph/0305161}}.

\bibitem{Qiu:2002mh}
J.-w. Qiu, {\em Nucl.Phys.} {\bf A715} (2003) 309--318,
\href{http://www.arXiv.org/abs/nucl-th/0211086}{{\tt nucl-th/0211086}}.

\bibitem{Martin:2009iq}
A.~Martin, W.~Stirling, R.~Thorne, and G.~Watt, {\em Eur.Phys.J.} {\bf C63}
  (2009) 189--285,
\href{http://www.arXiv.org/abs/0901.0002}{{\tt 0901.0002}}.

\bibitem{Gao:2013xoa}
J.~Gao, {\em et al.}, {\em Phys.Rev.} {\bf D89} (2014), no.~3 033009,
\href{http://www.arXiv.org/abs/1302.6246}{{\tt 1302.6246}}.

\bibitem{Ball:2012cx}
R.~D. Ball, {\em et al.}, {\em Nucl.Phys.} {\bf B867} (2013) 244--289,
\href{http://www.arXiv.org/abs/1207.1303}{{\tt 1207.1303}}.

\bibitem{Alekhin:2013nda}
S.~Alekhin, J.~Bluemlein, and S.~Moch, {\em Phys.Rev.} {\bf D89} (2014) 054028,
\href{http://www.arXiv.org/abs/1310.3059}{{\tt 1310.3059}}.

\bibitem{Owens:2012bv}
J.~Owens, A.~Accardi, and W.~Melnitchouk, {\em Phys.Rev.} {\bf D87} (2013),
  no.~9 094012,
\href{http://www.arXiv.org/abs/1212.1702}{{\tt 1212.1702}}.

\bibitem{Hirai:2007sx}
M.~Hirai, S.~Kumano, and T.-H. Nagai, {\em Phys.Rev.} {\bf C76} (2007) 065207,
\href{http://www.arXiv.org/abs/0709.3038}{{\tt 0709.3038}}.

\bibitem{Eskola:2009uj}
K.~Eskola, H.~Paukkunen, and C.~Salgado, {\em JHEP} {\bf 0904} (2009) 065,
\href{http://www.arXiv.org/abs/0902.4154}{{\tt 0902.4154}}.

\bibitem{deFlorian:2011fp}
D.~de~Florian, R.~Sassot, P.~Zurita, and M.~Stratmann, {\em Phys.Rev.} {\bf
  D85} (2012) 074028,
\href{http://www.arXiv.org/abs/1112.6324}{{\tt 1112.6324}}.

\bibitem{Bari:1985ga}
{BCDMS} Collaboration, G.~Bari {\em et al.}, {\em Phys. Lett.} {\bf B163}
  (1985)
282.

\bibitem{Benvenuti:1987az}
{BCDMS} Collaboration, A.~C. Benvenuti {\em et al.}, {\em Phys. Lett.} {\bf
  B189} (1987)
483.

\bibitem{Bodek:1983ec}
A.~Bodek {\em et al.}, {\em Phys. Rev. Lett.} {\bf 51} (1983)
534.

\bibitem{Gomez:1993ri}
J.~Gomez {\em et al.}, {\em Phys. Rev.} {\bf D49} (1994)
4348--4372.

\bibitem{Dasu:1993vk}
S.~Dasu {\em et al.}, {\em Phys. Rev.} {\bf D49} (1994)
5641--5670.

\bibitem{Adler:2006wg}
{PHENIX} Collaboration, S.~Adler {\em et al.}, {\em Phys.Rev.Lett.} {\bf 98}
  (2007) 172302,
\href{http://www.arXiv.org/abs/nucl-ex/0610036}{{\tt nucl-ex/0610036}}.

\bibitem{Abelev:2009hx}
{STAR} Collaboration, B.~Abelev {\em et al.}, {\em Phys.Rev.} {\bf C81} (2010)
  064904,
\href{http://www.arXiv.org/abs/0912.3838}{{\tt 0912.3838}}.

\bibitem{Khachatryan:2015hha}
{CMS Collaboration} Collaboration, V.~Khachatryan {\em et al.},
\href{http://www.arXiv.org/abs/1503.05825}{{\tt 1503.05825}}.

\bibitem{Buckley:2014ana}
A.~Buckley, {\em et al.}, {\em Eur.Phys.J.} {\bf C75} (2015), no.~3 132,
\href{http://www.arXiv.org/abs/1412.7420}{{\tt 1412.7420}}.

\end{thebibliography}\endgroup

\end{document}